# AN IR SEARCH FOR EXTINGUISHED SUPERNOVAE IN STARBURST GALAXIES

Bruce Grossan[1], Earl Spillar[2], Robert Tripp[3], Norbert Pirzkal[4], Brian M. Sutin[5], Paul Johnson[2]

[1]University of California at Berkeley Space Science Laboratory, Lawrence Berkeley National Laboratory Institute for Nuclear and Particle Astrophysics (INPA)

[2]University of Wyoming Physics and Astronomy Department

[3]Lawrence Berkeley National Laboratory Physics Division

[4]European Southern Observatory

[5]University of California Observatories/Lick Observatory, Santa Cruz, CA 95064, sutin@ucolick.org

## ABSTRACT

IR and Radio band observations of heavily extinguished regions in starburst galaxies suggest a very high SN rate associated with such regions. Optically measured supernova (SN) rates may therefore underestimate the total SN rate by factors of up to 10, due to the very high extinction ($A_B$ ~ 10-20 mag) to core-collapse SNe in starburst regions. The IR/radio SN rates come from a variety of indirect means, however, which suffer from model dependence and other problems.

We describe a direct measurement of the SN rate from a regular patrol of starburst galaxies done with K' band imaging to minimize the effects of extinction. A collection of K' band measurements of core-collapse SNe near maximum light is presented. Such measurements (excluding 1987a) are not well reported in the literature. Results of a preliminary K' band search using the MIRC camera at the Wyoming IR Observatory (WIRO), and an improved search strategy using the new ORCA optics are described. A monthly patrol of a sample of IRAS bright (mostly starburst) galaxies within 25 Mpc should yield 1.6 – 9.6 SNe per year, corresponding to the range of estimated SN rates. Our initial MIRC search with low-resolution (2.2" pixels) failed to find extinguished SNe in the IRAS galaxies, limiting the SN rate outside the nucleus (at > 15" radius) to less than 3.8 Supernova Rate Units (SRU or SNe per century per $10^{10}$ $L_{sol}$) at 90% confidence. The MIRC camera had insufficient resolution to search nuclear starburst regions, where starburst and SN activity is concentrated, explaining why we found no heavily obscured SNe. We conclude that high-resolution nuclear SN searches in starburst galaxies with small fields are more productive than low resolution, large-field searches, even for our sample of large (often several arc minutes) galaxies. With our ORCA high-resolution optics, we could limit the total SN rate to < 1.3 SRU at 90% confidence in 3 years of observations, lower than the most pessimistic estimate.

## AN IR SEARCH FOR EXTINGUISHED SNE IN STARBURST GALAXIES

### I. INTRODUCTION

Limitations of Optical Supernova Searches

A number of efforts to measure the frequency of supernovae (SNe) have been made which include systematic recording of observation times and sensitivities, and a relatively well-defined sample of galaxies (e.g. Muller et al. 1992, Evans, Van Den Bergh, & McClure 1989, Capallero & Turatto 1988). These works have all used optical observations, usually in B. Such observations cannot detect SNe behind significant amounts of dust, such as those occurring in dusty star formation regions, because the dust causes up to tens of magnitudes of extinction in the optical bands. It is difficult to correct optical rate measurements for this effect, since some estimates (discussed below) suggest that SNe in these regions could dominate the total rate. For this reason, SN rate measurements from optical observations can only be lower limits to the true rate. The SN rate from optical measurements (See Table I) is 0.7-1.6 supernova rate units (SRU), where 1 SRU = 1 SN century$^{-1}$ ($10^{10}$ $L_{sol}$)$^{-1}$, and $L_{sol}$ is the unit of solar luminosity. (Capallero & Turatto 1988, for example, refer to this unit as the SNu, but this term can be confused with the Solar Neutrino Unit or SNU from, e.g., Bachall 1989.) A rough estimate of the number of SNe per century in a galaxy similar to our own is ~ 1 SRU. The rate of core-collapse SNe (non-type Ia SNe) dominates the total rate measurement in late-type galaxies (e.g., Muller et al. 1992).

The Extinguished SN Hypothesis

Basic stellar evolution predicts that massive stars ( 8 $M_{sol}$; Woosley & Weaver 1986) burn their nuclear fuel rapidly and end their lives in a core-collapse SN in a short time ( about $10^{7.1\text{-}7.5}$ Myr. for stars of initially 8-15 $M_{sol}$ ; Schaller 1992). The progenitors of these core-collapse SNe are therefore found primarily in active star-forming regions, as they will explode before the burst is over. (The 1/e time of the star formation in the burst is ~ $10^{7.3-8}$ years; Rieke et al. 1980). The extinctions to stars in the largest star-forming regions, those in "starburst" galaxies, are measured to be ~ 14 - 25 mag (as measured for NGC 253 and M82, respectively; Rieke et al. 1980) in visual bands. Enormous far-IR luminosities are observed for starburst galaxies due to dust absorption and re-emission of light from the young, massive stars. Accounting for the observed spectrum shows that these highly luminous starburst galaxies harbor the largest populations of massive SN progenitor stars, particularly in their circumnuclear regions (Rieke et al. 1980). The *majority* of supernova progenitors are therefore undetectable with optical measurements in this simple hypothesis. Even a very recent optically-based measurment of the SN rate, specifically in starburst galaxies, found a rate similar to that measured in "normal" galaxies (0.7 $h^2$ SRU for Type Ib/c and ~0.6 $h^2$ SRU for Type II; Richmond, Filippenko & Galisky 1998). Optical measurements may therefore be significantly underestimating the true rate of SNe (e.g., Van Buren & Norman 1989).

The SN rate is important for understanding nucleosynthesis, the abundance of metals, and the structure of the interstellar medium. It is also important in the study of SNe themselves. Observations of SNe by gravitational wave and neutrino detectors would give us important new tools with which to study this phenomenon. To build such devices for the observation of SNe, however, one first needs to know the frequency of nearby SNe, including those that may be extinguished. In this paper, we describe a search for extinguished SNe using observations in the near-IR to overcome detection bias due to extinction. We present our observing strategy, expected event rate, prelminary results, and plans for an improved search.

## II. EVIDENCE FOR EXTINGUISHED SUPERNOVAE

One can observe dusty regions in the radio and IR because the resulting extinction is small in these bands. The SN rate estimates based on such observations are summarized in the second part of Table I, and range from 1.7-10 SRU. Radio observations of the two nearby starburst galaxies M82 and NGC 253 yield large numbers of bright compact sources consistent with young supernova remnants (SNRs; Kronberg & Wilkinson 1975, Antonucci & Ulvestad 1988), and more recently in M83 (Cowan et al. 1994). Unfortunately, to unequivocally prove that a given source is a young SNR, one must detect its slow fading. Typically, such observations require a baseline of a decade or more, and the required measurements are difficult due to the crowding of sources in galaxy nuclei. Kronberg & Sramek (1985) shows that some of the sources are fading, providing convincing proof that the sources are SN remnants, but Ulvestad & Antonucci (1994) disputes the findings. The latter work suggests that some of the sources may be flat spectrum sources (i.e., not SN related). Muxlow et al. (1994) confirmed that most sources are SNRs, but derived a low SN rate based on an assumed expansion velocity. Van Buren & Greenhouse (1994) uses a method of identifying SNRs and bases a SN rate estimate on the age of the full population of SNe, which is claimed to be a more direct method of measurement.

Rieke et al. (1980) used IR imaging and spectroscopy to produce a model of the stellar populations in M82 and NGC 253. In order for the model to be consistent with published radio, IR, UV, and X-ray data, an initial mass function heavily biased toward massive, young stars was required. A very high rate of extinguished SNe was therefore derived. The model-dependence of this method may be the reason why it yields higher rates than the most recent radio-based results.

The observations clearly indicate that enhanced star-formation is occurring in these starburst galaxies, and that there are large numbers of young, high-mass stars. The problem is that all of the SN rate estimates above have some model input and dependence. There can be little doubt that SNe do occur in extinguished starburst regions, however. A SNe was discovered in the IR K (2.2 μm) band in the super-IR luminous galaxy NGC 3690 (SN1992bu, Van Buren et al. 1994), i.e., right where one would expect such a detection. No optical detections of the SN were made in this frequently observed galaxy, consistent with heavy extinction to the SN. Unfortunately, no meaningful rate measurement can be derived from one event. The Extinguished SN hypothesis is therefore still without a robust confirmation or refutation, hence the motivation for our near-IR search.

## III. IR LUMINOSITY OF CORE-COLLAPSE SNE

To determine the required sensitivity for our search, we need to know the average near-IR luminosity of core-collapse SNe. Unfortunately, there is a paucity of published near-IR measurements of core-collapse SNe near peak brightness. Most published near-IR measurements are of Type Ia events. Since the possibility of host galaxy reddening plagues optical measurments, we have chosen not to extrapolate these data to determine the luminosity. Instead, we have sought near-IR measurements from the literature, and have added some of our own.

Detailed measurements of core-collapse SNe in the literature are dominated by studies of the nearby SN 1987a. This Type II is anomalous by many measures, including a >100 day wide light curve peak in the optical and near-IR bands, making the data on this SN of little value to us. This is *not* a case of, "any object close enough to be well-studied is anomalous"; the latter feature would be observed even if the SN were near the limits of detectability.

We were able to obtain near-IR measurements for 5 other core-collapse SNe, all except for one within a few days of maximum at K, from our own observations and from published measurements. These values are given in Table II. Core-collapse SNe include Type Ib and c, which are relatively similar, and the various Type IIs, which have a large dispersion in their optical luminosities (Miller & Branch 1990, hereafter MB). The optical Type Ib+Ic rate is 1.1 $h^2$ SRU and the optical Type II rate is about 1.6 $h^2$ SRU for spiral galaxies (Muller et al. 1992). It is therefore important to know the luminosity of both core collapse Type I and Type IIs. Since the K' band data of Panagia et al. 1986 sample a Type I core-collapse light curve fairly well near maximum light, we used these data as a light curve template to estimate the maximum brightness of other Type I core-collapse light curves with limited sampling. From our knowledge of the time of $B_{max}$ (maximum flux in the B band), we selected the measurement closest to the expected date of $K_{max}$ in order to make the smallest correction. We extrapolated to $K_{max}$ using the same magnitude changes per day relative to $B_{max}$ as in the Panagia et al. template. This one extrapolation in the table is small, making the SN only 0.14 mag brighter at our extrapolated maximum than in our actual measurements. We did not apply corrections for Type II light curves.

We found the maximum K-band luminosities to be $M_{kmax} = -18.0$ for Type Ibc SNe (the average of 1994I measured 15 days after optical max, 1983n at its $K_{max}$, and SN1984L corrected to $K_{max}$) and $M_{kmax}$ = –18.3 for Type II SNe (the average of 1993J at V max + 12 days and 1980k at $B_{max}$). (We use $H_o$ = 75 km $s^{-1}$ $Mpc^{-1}$.) Our data are probably slightly fainter than the true max, however, as the Type I core-collapse light curve of Panagia et al. might have missed $K_{max}$ by a few days, and for Type IIs, we have no good way to correct the data to $K_{max}$. It should also be noted that optical measurements suggest that there is a large intrinsic dispersion in the luminosity of our SNe (see below). With so few measurements, our values are therefore only a rough estimate of the average luminosities. While we recognize the limitations of these estimates, note that they are much more appropriate than theoretical values or extrapolations from optical measurements for our purposes. We adopt a combined value of $M_{kmax}$ = –18.1 for all core-collapse SNe in the absence of data that justifies subdividing them. Finally, note that the effect of extinction in K band must be insignificant here. The SNe used in the estimate above were all discovered optically, and could not have been discovered if more than a few mag of optical extinction were present. For a few mag extintion in the optical, the effect at K cannot be larger than a few tenths of mag.

In contrast to the near IR measurements, many SN light curves have been measured optically. Type Ibc have a blue luminosity of $M_{Bmax}$ = –17.1 with $\sigma$ =0.34 dispersion, whereas Type II have $M_{Bmax}$ = –16.9 with the large dispersion of 1.35 mag (MB, $H_0$ = 75 km/s/Mpc). (Note that Type Ibc is more luminous than Type II in MB, but conversely for our small number of IR measurements. This is consistent with the large dispersion in the luminosity of Type IIs.) Until more IR measurements are made, it will remain unclear whether this dispersion comes from variations in host environment extinction or whether it comes from intrinsic variations in SN luminosity.

## IV. OBSERVING PLAN & PREDICTED DETECTION RATE

### The Search Strategy - A Patrol of IRAS Galaxies in the K Band

Now, using our luminosity value derived above, consider observations of SNe in high-extinction starburst regions. Assuming 2 magnitudes of extinction at K (about 20 mag at B), the average core-collapse SN would have K=15.9 at a distance of about 25 Mpc (distance modulus = 32.0). Detecting a point source of this brightness is a modest task for modern detectors and IR optimized telescopes. A more critical factor is whether or not there are enough SNe within the volume one can search to make a meaningful measurement of the SN rate.

Massive stars in star-forming regions warm their enshrouding dust, and so most of their radiation escapes in the far-IR and is detected in the IRAS 60 and 100 μm bands. The IRAS far-IR luminosity ($L_{FIR}$; Soifer et al. 1987) of a galaxy is then a measure of its population of massive SN progenitors (unless other unusual sources of far-IR radiation dominate, such as a Seyfert nucleus). The SN rate should therefore be closely related to $L_{FIR}$, which is unaffected by extinction. (Type Ia SNe come from evolved stars in binary systems, and are not related to these massive stars, but these events are rare compared to core-collapse SNe and would not affect the total rate much; e.g. Muller et al. 1992). Kronberg et al. (1985) showed that radio flux from compact sources in starburst nuclei (probably SNRs) correlates with 100 μm flux, supporting this hypothesis. Using $L_{FIR}$, we can then predict the intrinsic (unextinguished) SN rates for dusty starburst galaxies. We assume the expected number of observed SNe for a given galaxy is the SN rate per luminosity times the $L_{FIR}$ for the galaxy multiplied by the cumulative visibility time. A single observation's visibility time is the lesser of the time until the next observation and the time for which a SN would be detectable at the distance of the given galaxy. We consider a sample of 177 IRAS catalog galaxies within 25 Mpc, which are visible from our northern hemisphere observatory ($\delta > -20$), for a measurement of the local SN rate. Core-collapse SN K band light curves have a fairly broad peak and typically stay within ~ 0.3 mag of peak for one month (e.g. Panagia et al. 1986, Elias et al. 1983, Dwek et al. 1983, Suntzeff & Bouchet 1991), allowing us to efficiently observe only once per month. Using the range of 1.7-10 SRU given above, and considering when during the year one can observe a given galaxy, 1.6-9.6 SNe/year should be observed within our 25 Mpc sample. (In those cases where $L_{FIR}$ was not given in Soifer et al. 1987, we generated these from the distances in Table 3 and the fluxes given in the IRAS faint source catalog.)

## V. OBSERVATIONS OF NEARBY STARBURST GALAXIES

Between 1992 and 1994, we performed observations on the 2.3 m Wyoming IR Observatory (WIRO) telescope in the K' band. (The K' band is centered at 2.1 μm, about 0.1 μm short of the K band center, sampling nearly the same band but with less thermal instrument background.) The best instrument then available was the MIRC (Michigan IR camera), with a 128 X 128 HgCdTl detector. At prime focus the instrument had a 2.2" pixel size. We were able to patrol more than 75 galaxies per night under good conditions, and we were granted 5 nights per month most months.

Although the background can vary with the outside temperature, we typically measured $\sigma = 17.8$ mag pix$^{-1}$ in sky-subtracted 120 s exposures. Therefore, our canonical K'=15.9 mag SN at 25 Mpc with 2 mag extinction in K' could be detected at 5 $\sigma$ in roughly 180 seconds in a single background subtracted image, assuming the signal is split between 2 pixels. (The pixel size is so much larger than the seeing disk here that additional pixels will only rarely contain significant flux). The read noise is negligible in exposures this long as the K' background is high.

The actual observing was conducted by taking three pairs of on- and off-source frames, each pair offset by ~11" in a different direction, all the same exposure time. Some short exposures off-source are also taken for use in making flats. Multiple, offset frames allow us to remove cosmic rays, to correct for bad pixels, and to rule out short-lived transient sources such as asteroids and satellites.

After our galaxy images are flattened and background subtracted, our software searches for SNe by registering each new image with a reference image and subtracting. This analysis is typically done on the same night as the new image is acquired to facilitate rapid follow-up observations. First, a median flat is made (typically from all the off-source observations during the previous nights of the run). Next, a background image is made by medianing the off-source observations. The flattened, background and dark-subtracted frames are then fed to the subtraction software. Because the 2.2" pixel size is large compared to the seeing, typically 0.7-1.0" FWHM, no seeing correction is required. The images are re-sampled to a finer scale (typically a factor of 5) to achieve sub-pixel shifting accuracy, then cross-correlated to find the optimum shift. A scaling factor is then applied by measuring a large aperture around the galaxy nucleus in each frame, so that the images have the same normalization. The images are then re-shifted and subtracted. The subtracted images are then searched for residual point sources, our SN candidates. We determine our total noise, including noise due to subtraction, directly from the subtracted images. We get good results in the disk of the galaxies, but large residuals in the bright, highly-peaked nuclei of galaxies. The results are discussed quantitatively in the next section.

As proof that our system was working, we detected 3 unextinguished SNe after their optical discovery (1993j, 1994d, and 1994i). In Fig 1, we show the subtracted images of SN 1994I. We measured the brightness of the SN to be K'=12.56 ± 0.040 on UT 1994 April 16. The noise was estimated using the variance of background counts for apertures placed around the subtracted image at the same galaxy radius as the SN. We calibrated the measurement with observations of HD 84800, using the measurements of Elias et al. (1982), and assuming negligible corrections between the K and K' bands. The standard observations were made seven hours before the observation of SN1994I, a potential source of error, but the sky was clear and stable throughout the night.

## VI. RESULTS & INTERPRETATION

Table III shows the number of galaxies imaged during each observing run. From this and the FIR luminosity of each galaxy, we can estimate that the expected number of observed SNe was 2.7-16 for our range of 1.7-10 SRU. No extinguished SNe were discovered. All of our detected SNe were easily detected in optical observations, and were therefore unextinguished. Normally, one would derive a 90% confidence upper limit of 3.8 SRU for the extinguished SNe rate in our 25 Mpc sample, but below, we show that we were not sensitive to SNe in all locations within our galaxies.

With the MIRC camera, we had a choice of observing at prime focus, where we would have an image scale of 2.2" per pixel and a field 4.7 arc min across, or (with some modification to the dewar) at Cassegrain focus with < 0.2" pixels and a field only 1% of the area at prime focus. We decided that the latter situation would be impractical, because of exposure time and because our most important galaxies, M82 and NGC 253, barely fit the frame at prime focus. We decided it would be a mistake to throw away so much of our field. However, what we gave up was the ability to search for SNe in the nuclear starburst region throughout most of our sample galaxies.

We could not search in the nuclear starburst region because our instrument had insufficient resolution. Joy, Lester, & Harvey (1987) describe the starburst region (as measured by CO emission) in M82 as ring-shaped with a peak-to-peak diameter of 450 pc. There is no reason to expect that larger central starburst regions are common in our nearby sample, so we assume the majority of our sample nuclear starburst regions are this size or smaller in the following argument. Consider the implications for our images of the core-collapse SN 1994I in M51 (8.3 Mpc away; Richmond et al. 1996; See Fig. 1) if it occurred in an extinguished nuclear starburst region. We measured 565 counts from this SN, in a 90 s image taken near maximum brightness. This optically detected SN was about 10 pixels from the galaxy center, where the subtraction noise was small (66 counts in our aperture or 18 counts/pixel). Scaling the M82 starburst region size by the relative distances of M51 and M82 yields a radius of the starburst region of about 3 pixels in M51. (Note that several works in this field adopt a distance of 3.2 Mpc for M82. We use the distance to its companion M81, 3.6 Mpc; Freedman et al. 1994.) Our subtractions are of poor quality in this central region. The standard deviation of an ensemble of apertures within a radius of 4 pixels from the galaxy center was about 140 counts. A comparison of this subtraction noise in the starburst region with the 560 count SN signal shows that we would only marginally detect ( 4 $\sigma$) starburst region SNe in M51. Since most of the galaxies in our (volume-weighted) 25 Mpc sample are much further, SNe in these galaxies would be even fainter, and therefore undetectable. The heavily extinguished SNe ($A_V$ = 10-20 mag) we are searching for would also have 1-2 mag of extinction even in our K' images, further reducing the signal to noise. This is why our detection rate fell short of the predicted range. The 2.2" pixels of the MIRC camera are too large compared to the ~ 0.7 - 1.0" FWHM seeing disk, so bright nuclei are too under-sampled to yield subtractions with low-noise at the nucleus, just where the SNe are predicted to be.

The productive results of this phase of our investigations are that we now have the experience and software for observing and reducing images of a large number of galaxies each night, and we have acquired the data to show that extinguished SNe, consistent with the hypothesis, must be concentrated in the nucleus (if present). A much smaller field is therefore acceptable, making future searches much easier with currently available instruments.

## VII. THE NEXT STEP: A HIGHER-RESOLUTION NUCLEAR SEARCH

In early 1995 we obtained the use of a NICMOS 256 $\times$ 256 array in the New Mexico IR Detector camera, which has much better noise characteristics, and discontinued observing with the MIRC. During this year, we designed and built field re-imaging optics, called the Optimum Resolution Camera Adapter (ORCA), to fully sample the point spread function (PSF). The camera dewar was enlarged to maintain these optics at cryogenic temperatures to minimize their thermal emission. This configuration yields a pixel size of 0.29" $pix^{-1}$, an appropriate match to our 0.7-1.0" seeing FWHM. Although our field is now only 1.2 arc min across, we are assured we will not miss a significant number of SNe as we already have the data to prove that they occur only rarely at galaxy radii outside the field.

Figure 2 shows an image of M82 with the ORCA. Note that we have demonstrated good background subtraction (no anomalous structure is evident in the image, the edges are smooth and faint, etc.) even though this large, nearby galaxy fills our field, and even though we need to point far from the galaxy for our background frames. In poor conditions we measure a background noise of 18.2 mag per pixel in 100 sec. on and off source, which permits us to measure a 15.8 mag star at 5 $\sigma$ in less than 900 sec. of integration. No data in typical conditions are available at this time. We also expect to increase our performance significantly by reducing the camera background through better alignment of the Lyot stop.

The higher resolution of the new camera gives us a very powerful tool for image subtraction- the ability to resolve the point spread function. By measuring the PSF of stars near a target galaxy, we can measure the PSF for each series of galaxy images even if there are no stars in our small field. During subtraction, the new or reference image, whichever has better seeing, will be smeared to match the PSF in the other image, yielding low subtraction residuals. This technique is commonly used in optical SN searches with good results. We have already implemented the software required to observe catalog stars for PSF measurement automatically during our galaxy observations.

A new search with the ORCA will produce a strict limit to the SN rate. Based on our 900 sec. integration time in poor conditions, adding read and slew times, taking 1/2 exposure time for the closer 1/2 of the sample, and a 66% acceptable weather factor, we should be able to observe 89 galaxies per typical 5 night (11 hour nights) observing run. (Due to longer exposures than with the previous camera, we will only repeat observations of galaxies with SN candidates.) Based on our projected values for camera performance in more typical conditions, and conservatively assuming one magnitude worse background than achieved for other cameras at the same site, we should be able to observe 142 galaxies per run (more than would be visible in our sample). Either way, we can observe the majority of our sample visible in any month to yield our expected 1.6 - 9.6 SNe per year. After 3 years of operation, the discovery of no SNe would limit the extinguished SN rate to < 1.3 SRU at 90% confidence, less than the most pessimistic non-optical estimate in Table I.

## VIII. SUMMARY

Optical measurements of the SN rate could be missing a significant fraction of SNe, particularly in the nuclei of starburst galaxies. Non-optical SN rates are, so far, derived from indirect measurements, not direct observations of SNe. These non-optical rates imply much higher rates that those given by optical searches, but they vary widely. Our direct, wide-field observations in the K' band show that in our IRAS galaxy sample, the SN rate must be < 3.8 SRU at 90% confidence outside galaxy nuclei (> 15" radius). This result allows future SN searches in nearby galaxies to cover only the nucleus of each galaxy (the inner ~450 pc). Because the galaxy nuclei are very bright, low-residual subtractions are required for such a search. From this requirement follows the need for adequate sampling of the instrumental point spread function. Our collaboration will proceed with such a search, with the required high-resolution camera, as soon as funding is available.

We note that a recent study of optical SN rates included an explicit measurement within the nuclei of starburst galaxies (Richmond, Filippenko & Galisky 1998). The authors of this work searched for changes in optical flux (Lick Observatory $R_S$ band) consistent with SN explosions within a fixed aperture containing the nucleus of each galaxy. Because of the bright nucleus background in the apertures, the sensitivity of this technique was poor and yielded a low cumulative visibility time. With no observed SNe, the authors were only able to limit the nuclear rate of unobscured SNe to < 12 $h^2$ SRU for Type Ib/c and < 7 $h^2$ SRU for Type II. As with other optical measurements, however, the study was insensitive to heavily obscured SNe. The unobscured SN rate in the very important nuclear region of starbursts therefore remains an open question of great interest.

The authors would like to thank Nino Panagia for making significant contributions of data and analysis to this paper. We also thank Harley Thronson for valuable scientific input, and generous donation of telescope time to our project. Special thanks are given to the faculty, students, and staff of the University of Wyoming Dept. of Physics and Astronomy in supporting Dr. Grossan during his visits, and to the staff of WIRO for their excellent technical support of this project. We acknowledge receipt of funding from the NASA GRO Guest Observer Program, the Lawrence Berkeley National Laboratory (LBNL) Physics Division, and the LBNL Institute for Nuclear and Particle Astrophysics.

## FIGURE CAPTIONS

Figure 1: K' Band images of NGC5194 (M51) with the MIRC camera on the WIRO Telescope. (a) shows the reference image, taken 1994 March 2. (b) shows the galaxy after the explosion of SN1994I, taken 1994 April 16. The SN is indicated by a light square around the object. (c) shows the results of registering, scaling and subtracting the 2 images. The residual SN is obvious, but significant noise is present at the location of the galaxy center. The image scale is 2.2"/pixel. The detector chip was not oriented parallel to standard compass directions. The SN is located 14"E and 12" S of the Nucleus (Richmond et al. 1996). Roughly speaking, north is at upper right and east is at lower right. A color version of these images is available in the Electronic Edition of the Astronomical Journal; another copy will be available at http://panisse.lbl.gov/public/bruce/irsn for at least several years after publication.

Figure 2: The ORCA K' image of NGC 3034 (M82) taken 1995 May 11 at the WIRO telescope. The image scale is 0.29"/pix. North is up, east is left. A color version of these images is available in the Electronic Edition of the Astronomical Journal; another copy will be available at http://panisse.lbl.gov/public/bruce/irsn for at least several years after publication.

**TABLE I**

**SN Rate Measurements**

| Rate (SRU) | Basis | Reference | Comment |
|---|---|---|---|
| | | **Optical Measurements** | |
| 0.74 | Optical | Capellaro & Turatto 1988 (Asiago Search) | |
| 1.43 | Optical | Evans et al. 1989 | |
| 1.56 | Optical | Muller et al. 1992 | Type Ibc dominate,most found in spirals. |
| ≈ 2 late-type (Sb - Im) ≈ 0.5 early-type (E-Sb) | Optical | Van den Bergh & Tammann 1991 | |
| | | **Unextinguished (Non-Optical) Measurements** | |
| 3 | radio | Kronberg & Wilkinson 1975 | 0.1 $yr^{-1}$ in M82 |
| 10 | Modeled IMF to match radio, IR, optical, X-ray observations. | Rieke et al.1980 | 0.3 $yr^{-1}$ in M82, same rate in NGC 253. Low mass stars must be suppressed. |
| 6-10 | Radio variability | Kronberg & Sramek 1985 | 0.2 - 0.3 $yr^{-1}$ in M82 |
| >3 | Radio Source Counts and Remnant ages in NGC 253 | Antonucci & Ulvestad 1988 | >0.1 $SNyr^{-1}$ in NGC 253 |
| 5 | "gas consumption rate" | Van Buren & Norman 1989 | 0.5 $yr^{-1}$ $10^{-11}$ $L_{sol}$ IRAS galaxy |
| < 3.4 | Age of Radio Souce Population and Source Counts (Used Log N-Log S arguments to throw away some Antonucci & Ulvestad sources) | Van Buren & Greenhouse 1994 | <0.10 $yr^{-1}$ in M82, <0.08 $yr^{-1}$ in NGC253 |
| <3.4 | Radio Variability | Ulvestad & Antonucci 1994 | <0.10 $yr^{-1}$ in M82, <0.25 $yr^{-1}$ in NGC253 |
| 1.7 | SNR number vs. Diameter relation | Muxlow et al. 1994 | 0.05 $yr^{-1}$ in M82 |

## TABLE II

## IR Measurements of Core-Collapse SNe

| SN (galaxy) | Spectral Type | Dist[1](Mpc) (ref.) | $M_{K\ max}$ | $K_{max}$ (date[2]; ref) |
|---|---|---|---|---|
| 1980K (NGC6946) | II-L | 5.5 (Buta) | -18.6 | 10.07 11/01, t($B_{max}$); (Dwek) |
| 1984L (NGC991) | Ib | 18.8 (Tully) | -18.6 | 12.73 at t($k_{max}$)[3] (Elias) |
| 1983N (M83 =NGC5236) | Ib | 4.7 (Tully) | -18.2 | 10.19 7/26.2 ≈t($K_{max}$); (Panagia ) |
| 1993J (M81 =NGC3031) | "IIb"[4] | 3.6 (Freedman) | -18.0 | 9.82 4/18.1= t($V_{max}$)+ 12, ~ $K_{max}$; (I5773) |
| 1994I (M51 =NGC5194) | Ic | 8.3 (Richmond) | -17.0 | 12.56 4/20= t($B_{max}$)+ 15 ~ $K_{max}$; (this work[5]) |

The average value of $M_{K\ max}$ for Type Ib and Ic is -18.0 ± 0.48
The average value of $M_{K\ max}$ for all Type II is -18.3 ± 0.34
The average value for all SNe together is $M_{k\ max}$ =-18.1 ± 0.30

Notes: Dwek = Dwek et al. 1983, Buta = Buta 1982, Tully = Tully 1988, Panagia = Panagia et al.1986, Freedman = Freedman et al. 1994, I5773 = Romanishin in IAUC 5773, Richmond = Richmond et al. 1996

[1]For all purposes in this paper, $H_0$ is taken to be 75 km $s^{-1}$ $Mpc^{-1}$.

[2]t($V_{max}$) or t($B_{max}$) + some number gives the time of maximum light in the given filter plus the given number of days. Note that the time of maximum K band brightness was ~10 days past $B_{max}$ in Panagia et al. 1986.

[3]Since the light curve for the Type Ib SN1983N was well sampled near max by Panagia et al. 1986, we used their data as a template to correct the observations of SN 1984L. The daily changes relative to B max were used, with linear interpolation, to correct the observations (K=12.87) to the maximum brightness. SN1994I was measured close to the predicted IR max, and so no correction was made. No corrections were made to Type II light curves as no representative light curve templates were available. See text for additional comments.

[4]A "IIb" refers to a SN Type II that exhibited Ib spectra at late time, with double maxima.

[5]These measurements were made at the University of Wyoming Observatory Telescope on UT 1994 April 16 with the Michigan IR Camera (MIRC). The measurements were made through a $K'$ filter, and the corrections to the K band were assumed to be small.

## TABLE III

## TABLE OF OBSERVATIONS

| Galaxy | D(Mpc) | Log($L_{FIR}$) | 93 Feb | 93 Mar | 93 Jul | 93 Oct | 93 Dec | 94 Feb | 94 Mar | 94 Apr | 94 May |
|---|---|---|---|---|---|---|---|---|---|---|---|
| NGC150 | 21.2 | 9.96 | | | | X | X | | | | |
| NGC157 | 22.0 | 10.35 | | | X | | X | | | | |
| NGC247 | 3.6 | 8.67 | | | X | m | X | | | | |
| NGC253 | 3.6 | 10.6 | | | | X | X | | | | |
| NGC278 | 11.8 | 9.62 | | | X | X | X | | | | |
| NGC337 | 22 | 10 | | | X | | | | | | |
| NGC578 | 22.6 | 9.81 | | | X | | | | | | |
| NGC598 | 0.8 | 9.11 | | | X | m | X | X | | | X |
| NGC613 | 19.8 | 10.31 | | | | X | X | | | | |
| NGC628 | 8.7 | 9.87 | | | m | X | X | X | | | |
| NGC660 | 11.5 | 10.38 | | | X | X | X | X | | | |
| NGC693 | 21.2 | 9.86 | | | X | | | | | | |
| NGC701 | 18.9 | 9.8 | | | X | | | | | | |
| NGC772 | 33.1 | 10.47 | | | X | | | | | | |
| NGC891 | 9.6 | 9.83 | | | X | X | X | | X | | |
| NGC908 | 17.8 | 10.23 | | | | X | X | | | | |
| NGC925 | 9.4 | 8.99 | | | X | m | m | X | | | |
| NGC972 | 20.6 | 10.5 | | | X | X | X | X | | | |
| NGC1022 | 18.5 | 10.18 | | | X | m | X | | | | |
| Maffei II | 5.0 | 9.72 | | | m | m | | X | X | | |
| NGC1055 | 12.6 | 10.08 | | | X | m | | X | | | |
| NGC1056 | 9.3 | 9.03 | | | X | m | X | | X | | |
| NGC1068 | 14.4 | 10.9 | | | X | | X | X | | | |
| NGC1084 | 17.1 | 10.34 | | | X | m | | X | | | |
| NGC1087 | 19.0 | 10.1 | | | | X | | | | | |
| NGC1156 | 5.0 | 8.54 | | | X | | X | | X | | |
| NGC1187 | 16.3 | 9.9 | | | | | | | | | |
| NGC1232 | 20.0 | 10.3 | | | | m | X | | | | |
| IC1953 | 22.1 | 10 | | | | | | | | | |
| NGC1385 | 17.5 | 10.1 | | | | m | X | | | | |
| NGC1415 | 17.7 | 9.7 | | | | X | | | | | |
| NGC1421 | 25.5 | 10.2 | | | | X | | | | | |
| IC342 | 4.7 | 10.06 | X | | X | X | X | X | X | | |
| UGC2855 | 15.4 | 10.34 | X | | X | X | X | X | X | | |
| UGC2866 | 16.4 | 10.38 | | | X | X | | X | X | X | |
| NGC1482 | 19.6 | 10.4 | | | | X | X | | | | |
| UGC2953 | 12.0 | 9.34 | | | X | X | m | | | | |
| NGC1530 | 32 | 10.35 | | | X | | | | | | |
| NGC1569 | 1.6 | 8.6 | | | X | X | X | | | | |
| NGC1637 | 8.9 | 9.13 | | | | m | X | | | | |
| UGC3190 | 20.8 | 9.74 | | | | | | | | | |
| NGC1832 | 25.8 | 10.05 | | | | | | | | | |
| NGC1964 | 22.3 | 9.99 | | | | | | | | | |
| 0553+03 | 10.3 | 9.19 | | | | | | | | | |
| NGC2139 | 24.5 | 9.99 | | | | m | | | | | |
| NGC2146 | 17.2 | 10.69 | m | m | | X | X | X | | X | |
| 0616-08 | 10.1 | 9.57 | | | | X | X | | | | |
| NGC2283 | 11.3 | 9.38 | | | | m | X | | | | |
| NGC2276 | 32.3 | 10.49 | | | | X | | | | | |
| NGC2403 | 4.2 | 9.31 | | m | | m | X | X | | X | |
| 0834-26 | 10.5 | 9.89 | | | | m | | | | X | |
| NGC2681 | 13.3 | 9.48 | | | | | | | | | X |
| NGC2748 | 20.0 | 10.2 | | | | m | X | | | | |
| NGC2798 | 23.4 | 10.59 | X | m | | X | X | X | | | |
| NGC2820 | 22.5 | 10.02 | | | | | | | | | |
| NGC2841 | 14.3 | 9.46 | | | | | | | | | |
| NGC2903 | 6.3 | 9.93 | X | X | | X | X | X | X | X | |
| NGC2964 | 21.9 | 10.2 | | X | | | | | | | |
| NGC2976 | 3.4 | 8.62 | | | | | | X | | | X |
| NGC2992 | 30.5 | 10.46 | | | | | | | | | X |
| NGC2985 | 17 | 10.1 | | | | | | | | | |
| NGC3044 | 20.6 | 10.06 | | X | | | | | | | |
| NGC3031 | 3.6 | 9.39 | X | X | | X | X | X | X | X | |
| NGC3034 | 3.6* | 10.45 | X | X | | X | | X | X | X | |
| NGC3067 | 24.2 | 10.2 | | | | | | | | | |
| NGC3079 | 20.4 | 10.63 | X | X | | X | X | | X | | X |
| NGC3077 | 3.4 | 8.6 | | | X | | | X | X | X | |
| NGC3166 | 22 | 9.93 | | | | | | | | | |
| NGC3169 | 19.7 | 9.98 | | | | | | | | | |
| NGC3147 | 37.6 | 10.74 | | | | | | | | | |
| NGC3177 | 21.1 | 10.1 | | X | | | | | | | |
| NGC3184 | 8.7 | 9.43 | | | | | | X | X | | X |
| NGC3198 | 10.8 | 9.52 | | | | | | | | | |
| NGC3227 | 20.6 | 9.95 | | | | | | | | | |
| NGC3294 | 26.7 | 10.21 | | X | | | | | | | |
| NGC3310 | 18.7 | 10.39 | | | | X | X | | X | X | |
| NGC3344 | 6.1 | 8.98 | | | | | | X | | | |
| NGC3351 | 8.1 | 9.42 | | | | X | | X | X | X | |
| NGC3353 | 16.8 | 9.57 | | | | | | | | | |
| NGC3368 | 8.1 | 9.45 | | | | X | | | | | |
| NGC3395/6 | 27.4 | 10.31 | | | | | | | | | |
| NGC3424 | 19.9 | 10.22 | | X | | | X | | | | X |
| NGC3432 | 10.9 | 9.32 | | | | | | | | | |
| NGC3437 | 17.2 | 10.23 | | | | | | X | | | X |
| NGC3448 | 18.4 | 9.96 | | | | | | | | | |
| NGC3486 | 7.4 | 9.06 | | | | | | X | | | |
| NGC3504 | 20.7 | 10.55 | X | | | X | | X | | X | |
| NGC3511 | 16.3 | 9.94 | | | | | | | | | |
| NGC3521 | 7.2 | 9.94 | | m | | X | X | X | X | | |
| NGC3556 | 14.1 | 10.3 | | | | X | X | | X | X | |
| NGC3583 | 28.5 | 10.42 | | | | | | | | | |
| NGC3593 | 5.5 | 9.16 | | | X | X | X | X | | | |
| NGC3623 | 13.5 | 9.22 | | | | X | | | | | |
| NGC3627 | 6.6 | 10.25 | X | X | X | X | X | X | X | X | |
| NGC3628 | 7.7 | 10.18 | X | X | X | X | X | X | X | X | |
| NGC3631 | 15.5 | 10.21 | | | X | | | X | | | |
| NGC3655 | 19.7 | 9.85 | | | | | | | | | |
| NGC3672 | 24.7 | 10.37 | | | | | | | X | X | X |
| NGC3675 | 12.8 | 9.86 | | | | | | | | | |
| NGC3683 | 22.1 | 10.49 | | | | X | X | | X | | |
| NGC3690 | 42.1 | 11.72 | X | X | X | X | X | | | | |
| NGC3726 | 16.7 | 9.87 | | | | | | | | | |

| | | | | | | | | | | |
|---|---|---|---|---|---|---|---|---|---|---|
| NGC3735 | 35.9 | 10.58 | | X | | | | | | |
| NGC3810 | 17.7 | 10.07 | | X | | | | | | |
| NGC3877 | 17.4 | 9.85 | | X | | | | | | |
| NGC3885 | 26 | 10.3 | | | | | | | | |
| NGC3887 | 15.1 | 9.8 | | | | | | | | |
| NGC3893 | 17 | 10.22 | | | | | | | X | |
| NGC3938 | 17 | 9.71 | | | | | | | | |
| NGC3949 | 17 | 9.89 | | | | | | | | |
| NGC3953 | 20.2 | 10.24 | | | | | | | | |
| NGC3955 | 17.9 | 10.01 | | | | | | | | |
| NGC3981 | 22.9 | 10.22 | | | | | | | | |
| NGC3982 | 14.8 | 9.98 | | | | | | | | |
| NGC4013 | 11.1 | 9.84 | | X | | | | | X | |
| NGC4027 | 22.4 | 10.4 | | | | | X | | X | |
| NGC4030 | 19.5 | 10.26 | | X | | | X | | X | X |
| NGC4038/9 | 20.8 | 10.81 | | | | | X | X | | |
| NGC4041 | 16.5 | 10.34 | | | | | | X | X | |
| NGC4045 | 26.6 | 10.28 | | | | | | | | |
| NGC4051 | 9.5 | 9.53 | | X | | | X | | X | |
| NGC4085 | 10.0 | 9.61 | | X | | | X | | X | |
| NGC4088 | 10.2 | 10.19 | X | X | X | X | X | | X | X |
| NGC4096 | 8.8 | 9.49 | | X | | | X | | X | X |
| NGC4100 | 14.3 | 10.08 | | | | | | | | |
| NGC4102 | 11.5 | 10.52 | | X | X | X | X | X | | |
| NGC4123 | 18.0 | 9.68 | | | | | | | | |
| NGC4157 | 10.3 | 10.11 | | X | X | X | X | X | X | |
| NGC4192 | 12.4 | 9.68 | | X | | | | | | |
| NGC4216 | 21.9 | 9.55 | | X | | | | | | |
| NGC4258 | 6.8 | 9.76 | | | | | X | | X | X |
| NGC4298 | 14.9 | 9.85 | | X | | | | | | |
| NGC4303 | 20.9 | 10.85 | | X | | X | X | | X | |
| NGC4321 | 21.0 | 10.33 | | X | | X | X | | X | |
| NGC4332 | 37.9 | 10.57 | | X | | | | | | |
| NGC4369 | 14 | 9.88 | | X | | | | | | |
| NGC4383 | 22.6 | 9.77 | | X | | | | | | |
| NGC4388 | 18.8 | 9.94 | | | X | | | | | |
| NGC4402 | 3.2 | 9.81 | | | X | | X | X | | X |
| NGC4414 | 9.7 | 10.14 | | | X | | X | X | | X |
| NGC4418 | 27.3 | 11 | | | X | | | X | | |
| NGC4419 | 16.8 | 9.84 | | | X | | | | | |
| NGC4438 | 16.8 | 9.77 | | | | | | | | |
| NGC4490 | 7.8 | 9.8 | | | o | | | X | X | |
| NGC4526 | 16.8 | 9.8 | | | | | | | | X |
| NGC4527 | 23.2 | 10.87 | | | X | | X | X | | X |
| NGC4532 | 20.5 | 9.97 | | | | | | | | |
| NGC4535 | 16.1 | 9.9 | | | | | | | | |
| NGC4536 | 24.3 | 10.83 | | | X | | X | X | | |
| NGC4559 | 13 | 9.79 | | | | | | | | |
| NGC4565 | 13 | 10.01 | | | X | | | | | |
| NGC4569 | 11.5 | 9.67 | | | X | | | | | |
| NGC4594 | 15 | 10.12 | | | X | | | | | |
| NGC4631 | 8.2 | 10.09 | X | X | X | | X | X | X | X |
| NGC4651 | 21.2 | 9.93 | | | X | | | | | |
| NGC4654 | 16.2 | 10.07 | | | | | | | | |
| NGC4666 | 21.9 | 10.93 | | | X | | X | X | | X |
| NGC4691 | 23.8 | 10.28 | | | | | | | | X |
| NGC4710 | 15.0 | 9.73 | | | | | | | | |
| NGC4725 | 23.3 | 9.82 | | | | | | | | |
| NGC4736 | 4.3 | 9.52 | | | X | | X | X | X | |
| NGC4781 | 16.9 | 10.1 | | | | | | | | |
| NGC4808 | 16.1 | 9.7 | | | | | | | | |
| NGC4818 | 14.0 | 9.67 | | | | | | | | |
| NGC4826 | 5.5 | 9.24 | | | X | | X | X | | X |
| NGC4845 | 17.4 | 9.94 | | | X | | | | | |
| NGC4900 | 17.7 | 9.69 | | | | | | | | |
| NGC5005 | 12.7 | 10.5 | X | | X | X | X | X | X | X |
| NGC5033 | 11.7 | 10.4 | X | | X | X | X | X | X | X |
| NGC5055 | 6.6 | 10.01 | X | X | X | | | X | | |
| NGC5170 | 25.7 | 9.3 | | | | | | | | |
| NGC5194 | 8.3* | 10.49 | X | X | X | | | X | X | X |

| | | | | | | | | | | | |
|---|---|---|---|---|---|---|---|---|---|---|---|
| NGC5195 | 9.3 | 9.47 | | | X | | X | X | | | |
| NGC5236 | 9.9 | 10.77 | | | | | | X | X | | |
| NGC5248 | 15.4 | 10.22 | | | X | | X | X | | X | |
| NGC5457 | 8.1 | 10.32 | X | | X | | X | X | X | | |
| NGC5506 | 24.1 | 10.2 | | | | | | | | | |
| NGC5678 | 25.7 | 10.52 | | | X | | X | X | | X | |
| NGC5676 | 28.1 | 10.67 | | | X | | | X | X | | |
| NGC5690 | 23.3 | 10.24 | | | X | | | | | | |
| NGC5713 | 25.3 | 10.69 | | | X | | | X | | | |
| NGC5719 | 19.7 | 10.18 | | | X | | | | | | |
| NGC5775 | 22.3 | 10.69 | | | X | | X | X | | | |
| NGC5792 | 25.7 | 10.4 | | | X | | | | | | |
| NGC5866 | 9.0 | 9.66 | | | X | X | | X | | | X |
| NGC5861 | 25 | 10.38 | | | X | | | | | | |
| NGC5900 | 34 | 10.54 | | | X | X | | | | | |
| NGC5907 | 8.9 | 10.16 | X | X | X | X | X | X | | X | |
| NGC5915 | 31 | 10.48 | | | X | | | | | | |
| NGC5929 | 35 | 10.55 | | m | X | X | | | | | |
| NGC5937 | 33 | 10.61 | | | X | | | | | X | |
| NGC5953 | 26 | 10.46 | | X | X | m | | | | | |
| UGC9913 | 72.7 | 12.12 | | | X | m | | | | | |
| NGC5962 | 26 | 10.45 | | | X | m | | | | | |
| NGC6015 | 16.5 | 9.6 | | | X | | | | | | |
| NGC6181 | 32 | 10.57 | | | X | X | | | | | |
| NGC6217 | 23.9 | 10.22 | | | m | X | | | | | |
| NGC6503 | 6.1 | 8.97 | | | X | X | X | | | | |
| NGC6574 | 30 | 10.47 | | | X | X | | | | | |
| NGC6643 | 25.5 | 9.96 | | | X | | | | | | |
| NGC6764 | 31.7 | 10.18 | | | X | m | | | | | |
| NGC6814 | 21 | 9.74 | | | | m | | | | | |
| NGC6835 | 22.8 | 10.06 | | | X | X | X | | | | |
| NGC6946 | 5.5 | 9.97 | | X | X | X | X | | | | |
| NGC6951 | 24.1 | 10.06 | | | X | | X | | | | |
| NGC7331 | 14.3 | 10.19 | | | X | X | X | | | | |
| NGC7448 | 29 | 10.33 | | | X | X | X | | | | |

| | | | | | |
|---|---|---|---|---|---|
| NGC7465 | 26 | 10.03 | X | X | X |
| NGC7469 | 66 | 11.4 | m | X | X |
| NGC7479 | 32 | 10.55 | X | X | X |
| NGC7541 | 36 | 10.83 | X | X | X |
| NGC7625 | 22 | 10.12 | X | X | X |
| NGC7640 | 13.0 | 9.17 | X | X | X |
| NGC7673 | 45 | 10.39 | m | m | X |
| NGC7678 | 47 | 10.64 | X | m | X |
| NGC7771 | 58 | 11.2 | X | | X |
| NGC7714 | 37 | 10.51 | X | | m |

An "X" indicates a successful observation. An "m" indicates a marginal observation, usually due to poor transparency. Distances are taken from Soifer et al. 1987, and where not in Soifer et al., from Tully 1988. A "*" indicates a distance from another source, as explained in the text. $L_{FIR}$ are taken from Soifer et al. 1987, or where not present, they were generated from the given distance and the IRAS Galaxies Catalog (Fullmer & Lonsdale 1989).